\documentclass[12pt]{iopart}
\usepackage[utf8]{inputenc}
 \usepackage[bottom=3cm]{geometry}
\usepackage{xcolor}
\expandafter\let\csname equation*\endcsname=\relax 
\expandafter\let\csname endequation*\endcsname=\relax
\usepackage{hyperref}
\usepackage{mathtools}
 
\usepackage{amsmath}
\usepackage{amssymb}
\usepackage{amsthm}
\usepackage{mathrsfs}
\usepackage{dsfont}

\begin{document}

\title{Quantum effects across dynamical horizons}
\author{Cecilia Giavoni$^\dag$ and Marc Schneider$^\ddag$}
\address{$^\dag$ Arnold Sommerfeld Center for Theoretical Physics, Theresienstra{\ss}e 37, 80333 M\"unchen, Germany}
\ead{cecilia.giavoni@physik.uni-muenchen.de}
\address{$^\ddag$ Institute for Gravitation and the Cosmos, University Park, Pennsylvania 16802, USA}
\ead{mms94@psu.edu}

\begin{abstract}
We present a generalization of the Hawking effect
for dynamical trapping horizons by calculating the tunneling 
rate in the Hamilton-Jacobi formalism. It turns out that
 all horizons classified by Hayward are
subjected to thermal quantum effects. While 
 the Hawking effect for future outer
and past inner trapping horizons is given as 
a particle emission, we show that 
the Hawking effect for future inner and
past outer trapping horizons 
translates to an absorption. 
The universality of the treatment allows a natural transfer to the static case.
\end{abstract}
\submitto{\CQG}
%
%
%
%
%

\section{\label{sec:level1}Introduction}

In its original version, the Hawking effect
describes particle production by black holes \cite{hawk75}.
The most common illustration uses Hawking pairs, that is to say, close to the 
black hole, a pair of particles is created and one particle 
falls into the black hole while the other escapes to future
infinity, where it is measured.
Energy budget considerations lead to the conclusion that
the black hole will lose energy because of this process which 
results in a shrinking radius. Hence, 
the in-falling particle could
in principle be interpreted as a particle with 
negative energy tunneling inside the black hole \cite{hawk76}. Hawking particles will potentially
cause the black hole to disappear.
Those calculations have led to plenty of follow-up articles,
mostly discussing information loss due to 
the Hawking effect involving a vast variety of
physical and mathematical ideas about the fate of the information inside.
A central issue in all those considerations, however, is the
nature of the Hawking process.
Its occurrence presumes necessarily - but not sufficiently -  
the presence of a horizon which acts as a 
separation principle for the Hawking pair \cite{vis03,unrh}. This is because
horizons are boundaries between two 
space-time regions which are to some extent causally disjoint. In other words, they
provide a suitable gravitational separation between the two Hawking partners.

The connection between particle production and the presence of a 
horizon suggests that black holes are just one example where this 
process occurs. In principle,
any kind of horizon might be furnished with the Hawking effect. 
A direct consequence \cite{gibb77} is the
belief that horizons could
be seen as thermodynamic objects with a specific
temperature. In the case of black holes, this temperature is
defined through the spectrum of the emitted Hawking quanta.
Although thermodynamic quantities
 are characterized with respect to
an averaging process, the Hawking temperature is sharply
defined. Nevertheless, it fulfills the property of a temperature
by being semi-positive and obeying a law analogous to the
first law of thermodynamics \cite{bek73}.

Physical considerations demand to formulate all predictions 
such that they are in principle measurable. For the initial Hawking 
proposal, the observer needs to be infinitely far away and has to
wait infinitely long. There the black hole
is characterised by an \textit{event} horizon:
well-defined in the asymptotic regions and static. However, this setup is 
unsuitable for practical purposes
because the event horizon itself is not 
physically observable \cite{vis90} by local experiments.
Hawking and Ellis \cite{hawk73} have revisited 
the definition of horizons to
describe local and dynamical black holes. However, this idea also involves
the global assumption that the space-time is asymptotically flat which is a severe restriction
since our universe is not considered to fulfill this assumption.
Hayward \cite{hay94} generalized Hawking and Ellis's idea to formulate a 
quasi-local definition of dynamical horizons 
resulting in the definition of \textit{trapping} horizons which does not require 
asymptotic flatness. 
These are classified concerning their causal structure
by using the area change along ingoing and outgoing
light rays which yields four basic types of trapping horizons. The black hole horizon,
for example, is categorized as future outer trapping horizon
(for the definition cf. section \ref{2}, or \cite{hay94}). 
In short,
trapping horizons can be physically understood as the horizon a specific 
observer perceives by testing the spacetime with light-torches:
chosen a point in spacetime, the observer switches the torch on and measures
how lightrays evolve in the nearby.
Hayward's notion is then clearly local and time-dependent.
This is of paramount importance for dynamical spacetimes but 
also when we intend to study the Hawking effect itself. In general,
Hawking particles might cause the horizon to change and 
therefore trapping horizons provide a more realistic framework.

While Hawking's analysis incorporates  
the Bogolubov coefficients, we want to work in
the tunneling picture \cite{par00,med05,arz05,ban08, seno15} to study the
Hawking effect for dynamical horizons.
Especially, the Hamilton-Jacobi method 
\cite{vanzo} has been proven to be 
 a powerful tool, reproducing for static black holes that
the horizon emits a thermal spectrum of particles.
This method
supports a strict generalization beyond the black hole case and
gives a similar result for Hubble horizons in
expanding cosmologies \cite{vanzo}, that is, the
observer will detect a 
leakage of Hawking particles into the Hubble sphere.
Besides past inner (Hubble horizon type) and 
future outer (black hole type) trapping horizons, two other types emerge:
future inner and past outer trapping horizons.
Examples for future inner horizons occur in collapsing cosmologies with a big crunch
while for a past outer the most prominent representative is a 
white hole. Nevertheless, there are systems with more than
one horizon, e.g. electrically charged black holes, like the Reissner-Nordstr\"om solution,
have future inner and future outer horizons while cosmologies
dominated by stiff matter acquire past inner and past outer ones.

In the following we will present the outline of this article: the second section
provides first an introduction to
Hayward's horizon classification, then we restrict ourselves
to spherically symmetric cases in order to provide explanatory examples. The third section
reviews the tunneling picture in the Hamilton-Jacobi method, and the fourth section
starts by applying this to future outer (black hole-type) trapping horizons.
From these considerations, we will investigate quantum effects for 
the remaining classes of trapping horizons to find a generalization of 
the Hawking effect. 
While we observe particle emission by future outer and 
past inner trapping horizons, we find as a main result
that the Hawking effect for past outer
and future inner trapping 
horizons turns out to be an absorption. In the fifth section, we give a
summary and discuss implications of this work
as well as a possibility to extend our analysis to more general spacetimes.

\section{Dynamical horizons}\label{2}

The relevant geometrical objects in the setup are horizons, which are a boundary
between observable and unobservable events. 

A black hole, for example, is defined in an asymptotically flat spacetime through its
event horizon. The exact localisation of the event horizon's position, however, presumes the knowledge about the whole evolution of the spacetime. Therefore, the notion of an event horizon is purely teleological which turns
this concept practically useless for observations made by local experiments. If one wants to
locate the event horizon, one needs to solve the Cauchy problem for the full future development 
of a set of Cauchy data.

In order to perform realistic experiments (locally and in finite time),
we need to abandon the mathematically handy notion of
event horizons for a more operational definition of
horizons which eventually can be measured by local experiments. 
These requirements culminated in the 
notion of dynamical 
trapping horizons, which are quasi-local concepts based on the local evolution of lightrays 
involving only a limited amount of global properties 
while being in-principle detectable. Let us investigate the different definitions in detail before 
going into specific examples.

General relativity defines the causal structure of a space-time ($\mathcal{M},g$) with 
pseudo-Riemannian manifold $\mathcal{M}$ and metric tensor $g$ by introducing a 
variety of different sets which allow to define an \textit{event horizon}. We assume all manifolds
to be time-orientable such 
that the terms "future" and "past" can be assigned globally and 
unambiguously. To develop an intuition for the structure of event horizons,
we consider the case of a Schwarzschild black hole. Following \cite{hawk73}, 
the na\"ive idea that horizons are separating boundaries
can be recast into mathematically robuster terms by
introducing the concept of the causal past $J^-$.
The set $J^-(\mathscr{S},\mathcal{M})$ is the 
region of space-time which causally affects events on a compact spacelike or null set $\mathscr{S}$. Taking the 
asymptotically flat exterior region of a Schwarzschild black hole, we specify the observer to be at
future null-infinity $\mathscr{I}^+$ and look at its causal past. Then we can define the black hole
event horizon to be the boundary 
$\dot{J}^-(\mathscr{I}^+,\bar{\mathcal{M}})$ where $\bar{\mathcal{M}}$ 
is the closure of the (asymptotically simple) manifold $\mathcal{M}$. 
Analogously, a past event horizon can be defined as 
$\dot{J}^+(\mathscr{I}^-,\bar{\mathcal{M}})$ which describes the white hole
event horizon in Kruskal space-time. \\
In the formalism of Hawking and Ellis, a trapped region of a manifold $\mathcal{M}$ is the set $\mathcal{M}- 
\bar{J^-}(\mathscr{I}^+,\bar{\mathcal{M}})$ that is the portion of $\mathcal{M}$ which
cannot communicate with any observer at $\mathscr{I}^+$; intuitively all lightrays from events in the
trapped region are destined to stay inside. The Schwarzschild black-hole interior is such a region. 
Note, an analogous definition can be found by investigating
$J^+(\mathscr{I}^-,\bar{\mathcal{M}})$. In this procedure, the leftover portion
would correspond to a past- or anti-trapped region which describes,
 for example, the interior of an eternal
white hole. As already said, these definitions are non-local and teleological
since the observer rests at $\mathscr{I}^+$ or $\mathscr{I}^-$,
i.e. they require the knowledge
about the entire causal past/future $J^\pm$ and are non-dynamical.

A remedy was found by Hayward \cite{hay94} who constructed a general notion of 
dynamical horizons by introducing the double-null foliation:
 let $(\mathcal{M},g)$ be a four-dimensional,
globally time-orientable space-time which is foliated into spacelike hypersurfaces $\Sigma_t$ 
 along the temporal coordinate $t$. Then,
consider a 2-dimensional, compact, orientable,
space-like surface $\mathcal{S}$ on $\Sigma_{t}$ \cite{hay06} and two half-open intervals 
$\mathcal{I}^\pm=[0,I^\pm)$ such that we can define a smooth embedding 
$\iota:\mathcal{I}^+\times\mathcal{I}^-\times\mathcal{S}\to\mathcal{M}$. 
We can construct two null congruences $l^\pm$ (outgoing "+" and ingoing "--") 
which are orthogonal to $\mathcal{S}$ and allow for a definition of $\mathscr{L}_\pm$, the Lie
derivative with respect to a normal direction to $\mathcal{S}$. To formulate a quasilocal
description of a horizon, we will use the expansion 
$\theta^\pm=h^{-1}(\mathscr{L}_\pm h)$ of $\mathcal{S}$ 
along the null congruences $l_\pm$ which describes whether 
$\mathcal{S}$ will expand ($\theta^\pm>0$) or shrink ($\theta^\pm<0$)
if 
infinitesimally dragged along $l_\pm$ \footnote{Similar concepts were used by
Hawking and Ellis \cite{hawk73} to 
define apparent horizons. However, this description imposes two global assumptions, i.e. asymptotic 
flatness and regular predictability, absent in Hayward's notion. Asymptotic flatness, for example, 
cannot be realized for a de Sitter universe filled with a positive cosmological constant.}
; $h$ is the induced two-dimensional metric on $\mathcal{S}$.

By definition, a compact two-dimensional spatial surface $\mathcal{S}$ is
a trapped surface, if $\theta^+\theta^->0$ everywhere on 
$\mathcal{S}$. 
In particular, if both $\theta^{\pm}<0$ everywhere on $\mathcal{S}$ 
we call $\mathcal{S}$ a future trapped 
surface while if $\theta^{\pm}>0$ everywhere on
$\mathcal{S}$, then $\mathcal{S}$ is called a past-trapped surface, later 
referred to as anti-trapped surface.
The portion $ \mathscr{T}$ of $\Sigma_{t}$ that is foliated by
$\mbox{(anti-)}$trapped surfaces, is known as the \textit{(anti-)trapped
region} of $\Sigma_{t}$.
We have seen that with the help of light rays, we can study the causal structure
of spacetimes and detect trapped regions by a local experiment using torches.
In a flat spacetime, future-directed outgoing 
light rays diverge ($\theta^+>0$) while ingoing light rays converge ($\theta^-<0$). 
This behavior characterizes \textit{normal regions}.
 On the other hand, in a \textit{trapped region} $\mathscr{T}$
both, ingoing and outgoing light rays are converging, i.e. 
$\theta^{+}<0$ and $\theta^{-}<0$. In other words, 
everywhere in $\mathscr{T}$ the light-cone is so bent that all future-directed
signals are trapped in $\mathscr{T}$ and cannot escape.

Black holes can therefore be defined by their future trapped regions
because the light which has fallen inside is destined to
stay in the interior.
An expanding universe, in contrast, admits an 
anti-trapped region ($\theta^\pm>0$)
beyond the Hubble sphere with
radius $R=R_H$. In a homogeneous and isotropic slicing, the expansion will apparently reach
superluminal speed at $R>R_H$ \cite{dav04}, such that the effective
speed along the ingoing direction is not fast enough
to counteract the expansion, ending up in a net outgoing
motion. Indeed, the future light cone in the anti-trapped region is so bent towards the
outward direction that signals coming from sources outside ($R>R_H$) are excluded 
from the normal region inside ($R<R_H$).

A crucial concept in the definition of \textit{trapping horizons}
involve marginally trapped surfaces which
are spatial two-dimensional surfaces $\mathcal{S}$ on which either $\theta^+=0$ (future marginally trapped) or  
$\theta^-=0$ (past marginally trapped)\footnote{The definition in this article exchanges the null congruences $l_\pm$
for past and future which
deviates from the original definition in \cite{hay94} where $l_\pm$ are fixed. In the framework used in \cite{hay94},
$\theta^+=0$, while the distinction between future and past depends on the sign of $\theta^-$. Except for peculiar cases, e.g. plane-wave spacetimes,
both notions coincide (cf. \cite{faraoni} for more details).}. 
A \textit{future} trapping horizon is the 
closure $\bar{\mathcal{H}}$ of a three-dimensional 
surface $\mathcal{H}$ which is foliated by 
marginally trapped surfaces with the condition $\theta^+=0$
and $\theta^-<0$ and respectively $\theta^-=0$
and $\theta^+>0$ for \textit{past} trapping horizons. In other words,
the union of marginally trapped surfaces along the time flow constructs a dynamical notion
for horizons \cite{hay06}. It should be noted, that the occurrence of the trapping
horizon is independent of the chosen foliation \cite{vanzo}.

Dynamical horizons could be either space-like or time-like,
however, if for a considerable 
amount of time no matter has crossed, the horizon will stabilize
 and form a
static null horizon eventually approaching an event horizon \cite{ash01,ash02, ash03}
in asymptotically flat spacetimes.
 Therefore, the treatment of dynamical horizons clearly covers static horizons but it
 should be mentioned that trapping horizons can occur even in systems which never form an event
 horizon \cite{hay06}.
 
 We can study the behaviour of the null expansions
across the marginally trapped surface by involving the Lie derivative $\mathscr{L}_\mp\theta^\pm$ 
which shows in which direction specific lightrays (outgoing or ingoing) are trapped. We call a future 
horizon \textit{outer} if $\mathscr{L}_-\theta^+<0$ and \textit{inner} if $\mathscr{L}_-\theta^+>0$ 
at $\mathcal{H}$. For past horizons, we can define similarly an outer horizon as
$\mathscr{L}_+\theta^-<0$ and the corresponding inner horizon $\mathscr{L}_+\theta^->0$ to hold
at $\mathcal{H}$. Note, the expressions  $\theta^\pm\theta^\mp$ as well as
 $\mathscr{L}_\mp\theta^\pm$ are geometrical invariants in the sense of \cite{hay94}.

For illustrative reasons, we will change to the spherically
symmetric framework and discuss some examples. It should be noted that the areal radius $R$ is 
fully characterising the surface of symmetry.
Consider the horizon to be at $R_H$, the normal region can be either outside the horizon ($R_{\rm normal}>R_H$) like for a black hole or inside ($R_{\rm normal}<R_H$) 
like for a Hubble sphere in an expanding cosmology. Mathematically,
this property is reflected by the Lie derivative of the expansion 
$\theta^\pm$ along in- or outgoing light rays. 

Hence, trapping horizons can be classified into four types: future horizons
are defined using $\mathscr{L}_-$, the Lie derivative along ingoing 
light rays at the horizon, because $\theta^\pm<0$ in the trapped region indicates
that the classically allowed direction is ingoing, i.e. this is
the null geodesic unchanged by geometry:
\begin{itemize}
\item[(a)]\textbf{future outer trapping horizon } (FOTH): 
$\mathscr{L}_-\theta^+|_{R=R_H}<0$
\end{itemize}
FOTHs are the most prominent type since they cover
black hole horizons: the horizon
lies in the future of the observer who can cross it from the normal region,
outside at $R>R_H$, into the 
trapped region inside the horizon.
\begin{itemize}
\item[(b)]\textbf{future inner trapping horizon } (FITH): 
$\mathscr{L}_-\theta^+|_{R=R_H}>0$
\end{itemize}
Examples for FITHs are big crunch scenarios of
contracting cosmologies: the horizon is the surface where a homogeneous and 
isotropic collapse, 
as seen from an observer at $R=0$, is happening 
at the speed of light $c$. Beyond the horizon, the contraction 
appears to be faster than $c$ \cite{dav04}, hence, there is
no chance to escape the collapse. For FITHs, the 
normal region is inside, at $R<R_H$, and signals from the observer 
will remain inside the horizon. 

In contrast to future horizons, the Lie derivative
along outgoing light rays $\mathscr{L}_+$ is used
to characterize past horizons. Note that in this setup
the outgoing direction is classically favored:
\begin{itemize}
\item[(c)]\textbf{past inner trapping horizon } (PITH): 
$\mathscr{L}_+\theta^-|_{R=R_H}>0$
\end{itemize}
This type describes, for example, an observer in 
an expanding universe surrounded by a Hubble sphere.
The normal region is inside the Hubble horizon and signals from events beyond, 
i.e. at $R>R_{H}$,
will not enter the normal region (as long as 
the time evolution does not collect them naturally).
Indeed, in the 
observer's reference frame,
the expansion will push signals apart faster
than the speed of light \cite{dav04} at $R>R_H$.
Another notable example of PITHs is the horizon of the de Sitter spacetime \cite{gibb77,akh06}.
\begin{itemize}
\item[(d)]\textbf{past outer trapping horizon } (POTH): 
$\mathscr{L}_+\theta^-|_{R=R_H}<0$
\end{itemize}
An observer in the normal region at $R>R_H$, i.e. outside the horizon,
will localize the horizon in the past light cone.
Examples like white holes have the feature 
that radial geodesics lead away from 
this object and make it impossible to enter, 
consequently, all observers in the interior will be released into
the normal region.

Reissner-Nordstr\"om black holes are solutions of Einstein's equations where 
future inner and future outer 
horizons occur. If we assume they were suffering from an evaporation like
Schwarzschild black holes, we must 
incorporate the Hawking effect for future outer and future inner horizons;
even in the case of black hole formation, 
we can encounter the presence of both horizons (cf. \cite{hay06}) such that
a general description for (dynamical future) horizons is of paramount importance to
analyze gravitational collapses concisely.

In contrast to static spacetimes, dynamical spacetimes do not admit a 
global timelike Killing vector field. Therefore, it seems that there is no preferred 
time direction along which a notion of energy could be defined.
However, Kodama \cite{kod80} found that in spherically symmetric spacetimes exists 
a vector field $K$ which is divergence-free and generates a preferred time-flow
together with an associated energy flux \cite{ab10}. 
The Kodama vector is defined by \cite{hay09} $K=g^{-1}(\ast \mbox{d} R)$ with $\ast$ being the Hodge star in the space perpendicular to the spheres of symmetry, or in components
$K^a=\epsilon^{a b}_{\perp} \nabla_b R$ where $\epsilon^{a b}_{\perp} $ is the $(1+1)$-dimensional Levi-Civita tensor in the temporal-radial plane.
In particular, this definition involves that $\mathscr{L}_KR=0$ which
can be interpreted such that $K$ is always orthogonal
to the spheres of symmetry \cite{far17}. 
In the normal region, $K$ is time-like, on the horizon null, and in 
the trapped region space-like; for static spacetimes, $K$ is parallel
to the Killing vector field. 

By the use of the Clebsch decomposition in $(1+1)$-dimension, one can show that the Kodama vector 
$K$ naturally induces a temporal coordinate, the \textit{Kodama time t}, in the regions where
$K$ is timelike. Hence, $K$ defines a preferred class of fiducial 
observers which move along integral curves associated to $K$ with velocity
$V=K/\|K\|$ for which $K\propto\partial_t$ \cite{ab10}.
These properties of the Kodama vector allow to introduce an invariant (cf. \cite{acqua} for details)
surface gravity notion, namely the Hayward-Kodama surface gravity which is defined through
$K^\nu\nabla_{[\mu} K_{\nu]}=\pm\kappa_HK_\mu$ at $R=R_H$. Hayward \cite{hay94} 
provided an operational definition by
$\kappa_H=\tfrac{1}{2}\ast\mbox{d}\ast\mbox{d} R=\tfrac{1}{2}\Box_\gamma R$ 
at the horizon. The expression
$\ast\mbox{d}\ast$ is a divergence, and d the gradient which allows to rewrite
$\ast\mbox{d}\ast\mbox{d}=\Box_\gamma=\gamma^{-1}(\mbox{d},\mbox{d})$. The $\Box_\gamma$
operator is constructed with
respect to the metric $\gamma$ of the
two-dimensional space normal to the spheres of symmetry \cite{hay98}. 
This astronomical quantity describes for static black holes
the (gravitational)
acceleration experienced by a test particle at
the horizon. For generic spacetimes
exists no such interpretation \cite{niel08} and
the surface gravity can only be understood as a parameter connected to 
the temperature.

Additionally, a fiducial observer in the normal region, can exploit the proportionality
between $K$ and $\partial_t$ and deduce the invariant Kodama-energy along the flow of $K$, similar to the energy associated to the 
Killing vector in static spacetimes.  For a particle with action $S_0$, the Kodama energy is defined as $\omega=-K^\mu\nabla_\mu S_0$.
Although one can choose a Kodama foliation and refer to a fiducial observer, it should be stressed 
that this remains a choice. One can always refer to different local observers thanks to the individual
covariance of the energy $\omega$ and the surface gravity $\kappa_H$.

In the following, we will restrict ourselves to spherically symmetric 
spacetimes to illustrate how a generalisation of the Hawking effect can be achieved employing a 
certain symmetry.
Generally, a four-dimensional spherically symmetric spacetime can be locally coordinatised by the 
the metric $g=\gamma_{ij}(x)\mbox{d}x^i\mbox{d}x^j+R^2(x)\mbox{d}^2\Omega$ with signature $(-,+++)$ \cite{bine15}
where $\gamma$, defined as above, covers the temporal-radial plane and $R(x)$ depends on the 
$x^i$ explicitly. The solid angle is given by d$^2\Omega=\mbox{d}\vartheta\otimes\mbox{d}\vartheta+\sin^2(\vartheta)\mbox{d}\varphi\otimes\mbox{d}\varphi$.
Notice that in a Kodama foliation, corresponding to a Kodama observer,
any metric in the \textit{Kodama coordinates} $t$ and $R$
 would acquire a diagonal form \cite{far17}
\begin{equation}\label{kodmet}
g=g_{tt}(t,R)\mbox{d}t\otimes\mbox{d}t+g_{RR}(t,R)\mbox{d}R\otimes\mbox{d}R+R^2 \mbox{d}^2\Omega,
\end{equation}
due to the properties $\mathscr{L}_K R=0$ and $K\propto \partial_t$.
To properly describe quantum processes across the horizon, we refer to a coordinate system which 
covers both regions, trapped and normal and is regular at the horizon.
Therefore, the concrete calculations in this article will be carried out in the 
Eddington-Finkelstein-Bardeen metric \cite{bard73}
\begin{equation}\label{bardeen}
g=-e^{2\psi(\xi,R)}C(R)\mbox{d}\xi\otimes\mbox{d}\xi
\pm2e^{\psi(\xi,R)} \mbox{d}\xi\otimes\mbox{d}R
+R^2\mbox{d}^2\Omega
\end{equation}
with $\mbox{d}^2\Omega$ denoting the solid angle, $R$ the 
areal radius, and $C$ and $\psi$ functions which stipulate
the properties of the horizon. At the horizon $R=R_H$, 
the derivative of $C(R)$ is assumed to be finite and different from zero.
The variable $\xi=\{v,u\}$
covers the retarded or advanced light cone coordinates with $u=t-R^\ast$ being the 
outgoing, $v=t+R^\ast$ the ingoing light cone coordinate,
and $R^\ast=\int_R\tfrac{dR}{C(R)}$. 
One choice, $\xi=v$, is particularly used to treat 
future trapping horizons, since $v$ is well-behaving across the horizon,
that is to say, $v$ remains ingoing even in the trapped region;
for the same reason, the other choice, $\xi=u$, is best suitable for past trapping horizons. 
The off-diagonal term in \eqref{bardeen}
acquires a plus sign for $\xi=v$ and a minus sign for $\xi=u$.

Note, if at the horizon $\mathscr{L}_{\pm}\theta^\mp=0$ then the 
horizon is degenerate, i.e. it is an inner and outer horizon at the same time and 
$\kappa_H\equiv0$ \cite{hay98}.
This type occurs for example in a radiation dominated universe.

\section{Hamilton-Jacobi formalism}

In the previous section, we noticed that trapping
horizons could behave as semipermeable 
boundaries. Then, the Hawking effect is proposed to be the process 
acting \textit{against} the geometrically preferred direction of horizon
crossing; this heuristic idea will be
the guiding principle
throughout this article and the following calculations.
We will introduce the Hamilton-Jacobi tunneling method,
which enables us to 
study the change of the particle number in the presence of dynamical horizons through 
tunneling and provides an intuitive understanding of the 
ongoing effects. This picture was first suggested by
Parikh and Wilczek \cite{par00}, Massar and Parentani \cite{mass00}, and has been 
connected to the Hamilton-Jacobi formalism by Padmanabhan \cite{paddi},
and thoroughly reviewed by Vanzo et. al. \cite{ang05,di09,vanzo}
Basically, in the normal region, near $R_H$,
a particle-antiparticle pair is created due to 
the excitation of the vacuum by the strong gravitational field; 
the horizon itself is interpreted as sort of barrier \cite{par04}
inducing a pole in the propagator of the particle.
Once the pair is created,
the tunneling of one particle through the horizon could separate them,
preventing the pair to recombine. The remaining particle will
eventually propagate to the observer where it is measured as \textit{Hawking particle}.

In this article, the gedankenexperiment guiding
our analysis, independently of the horizon type, will be the following: in the normal region, where the Kodama vector is timelike, the Kodama-observer prepares a state
at an initial (Kodama-) time $t_{\rm start}$ and counts a certain number of particles.
Then the initial conditions are evolved to a later time 
$t_{\rm end} > t_{\rm start}$ and the observer will count again.
If the detection counts
more particles we will say the Hawking effect is an \textit{emission}
of particles by the horizon, if fewer particles are counted, we will interpret 
the Hawking effect as an \textit{absorption} of particles by the horizon. The particle count can be realised by a clicking event of an Unruh-DeWitt detector
\cite{acqua} which moves along the Kodama flow such that the energy of the particle can be
related to the invariant Kodama energy.

To implement these tunneling phenomena mathematically, the appropriate
description would be to find emission and absorption rates in 
quantum field theory on curved spacetimes. However,
a quantum mechanical formulation by a WKB approximation
and the quantum field theoretic description agree for quasi-local
states: the field-theoretical analog of a tunneling 
would be the propagation across
$R=R_H$, given by the two-point function $\langle\phi(R_1)\phi(R_2)\rangle$
with $R_1$ and $R_2$ on opposite sides with respect to the horizon. It should 
be mentioned, the choice
of the vacuum is crucial to define the energy spectrum seen by the observer. Since the 
number operator is state-dependent, we will argue that there is a special or preferred vacuum 
which describes the energy spectrum of the two-point
function seen by an observer which moves along trajectories generated by a vector field reflecting the
symmetries.  
In spherical symmetric spacetimes this is given by the Kodama vector and hence the choice of a Kodama foliation \cite{kod80}.
For quasi-local states,
the two-point function can be related to the WKB rate \cite{mor11,col14},
 which is an observable \cite{di09}
as well. Therefore, we are allowed to use techniques from the WKB framework of a
particle in a potential as long as our states are suitably localized.
The WKB approximation is valid in regimes where the
effective potential, induced by the geometry, 
varies slowly in time compared to the frequency of the particle.
This approach allows us to describe even slowly varying, time-dependent,
spacetimes.
To be consistent with the WKB condition, we 
assume the Kodama energy of the tunneling particles $\omega$ to be small
compared to the energy scale set by the classical geometric background which
in terms of \eqref{bardeen} translates to $\partial_RC(R)\ll 2\omega$. 

While the Hawking effect itself gives dynamics to the horizon,
the effect is, however, too small 
to violate the slow-evolution assumption set by the WKB approximation.  
For a black hole this can be justified by a back-of-the-envelope
calculation: consider a stellar black hole with mass $M\approx 10^{30}
\:\mbox{kg}$ and radius $r_s=2G_NM/c^2\approx 1500\:\mbox{m}$ where $c$ is the speed of 
light and $G_N$ Newton's constant. 
An emission of an extremely heavy particle with
almost Planck mass $M_P\approx10^{-8}\:\mbox{kg}$
changes the Schwarzschild radius and consequently the scale induced
by the geometry by $\delta r_s\approx10^{-35}\:\mbox{m}$ which
is negligible even in the case of numerous particle emissions.

The basic idea of the Hamilton-Jacobi method connects 
the tunneling probability 
to an imaginary contribution in the classical action of the particle $S_0$.
For our analysis we consider a scalar field $\phi$ with mass $m$ 
satisfying the Klein-Gordon equation
\begin{equation}
\left(\Box-\frac{m^2}{\hbar^{2}} \right)\phi=0
\label{kleingordon}
\end{equation}
where $\Box$ is constructed with respect to \eqref{bardeen}.
Note that the description of tunneling processes through black hole horizons 
is state-independent for scalar fields \cite{mor11}.
Within the validity of the WKB approximation,
we take
\begin{equation}
\phi= e^{\tfrac{i}{\hbar}S}=e^{\tfrac{i}{\hbar}S_0+S_1+\mathcal{O}(\hbar)},
\label{eis}
\end{equation}
as ansatz for the solution to \eqref{kleingordon},
where we included $\hbar$ explicitly which serves
as smallness parameter in the second step. Furthermore, we expanded the complex 
action $S$ to incorporate quantum effects to first order in $\hbar$.
After substituting this approximation into \eqref{kleingordon} and splitting the resulting 
equation into real and imaginary part,
we take the semi-classical limit, $\hbar\to0$, and obtain the relativistic
 Hamilton-Jacobi equation for the classical action
\begin{equation}\label{HJ}
g^{-1}(\mbox{d}S_{0},\mbox{d}S_{0})+m^{2}= 0.
\end{equation}
In \cite{vanzo}, it has been argued that the tunneling paths are
most likely given by null-like trajectories and the mass term in the Hamilton-Jacobi equation 
can be neglected. Even if considering timelike trajectories one can show that 
the result does not substantially deviate from the massless case \cite{ang05}.
Taking this into account, the
above equation can be solved with the following ansatz for $S_{0}$:
\begin{equation}\label{phiwkb}
S_{0} =\int \partial_\xi S_0 \mbox{d}\xi+\int \partial_RS_0 \mbox{d}R.
\end{equation}
We perform the integration along the dynamical path of the particle but we should be aware
that we could in 
principle encounter poles which can be avoided by a complexification of the 
manifold, resulting in an imaginary part in the classical action $S_0$. All
integration ranges will be specified when we analyse the specific horizons.
The $\xi$-integral is vanishing along null tunneling paths and therefore, does not contribute to 
Im$(S_0)$ \cite{vanzo}.
Instead, the horizon induces a pole in the radial integration 
at $R_H$ (see next section or \cite{vanzo1}), which exactly describes the 
classically forbidden path.
We suppressed the angular part because, there will be no pole in the $\vartheta$- and 
$\varphi$-integration due to the underlying 
spherical symmetry \cite{sar08}.
Analogous to quantum mechanics, where tunneling processes through a barrier result from 
imaginary momenta, gravitational tunneling is facilitated by an
imaginary contribution to $S_0$. The WKB wave-function experiences a discontinuity
at the pole but will be able to pass around on a complex path. Within this framework,
the rate for an individual particle to tunnel is given by
 \begin{equation}\label{gamma}
 \Gamma \propto e^{-\tfrac{2}{\hbar}{\rm Im}(S_0)},
 \end{equation}
which shows the immediate relation to the imaginary part of the 
classical action \cite{van11}. Expression \eqref{gamma} describes always
the classically forbidden process whenever the rate is exponentially suppressed.
This is the case for gravitational tunneling, hence, we claim for all
realizations of the Hawking effect that Im$(S_0)>0$. Furthermore, 
the rate in \eqref{gamma} adopts the exponent's diffeomorphism invariance which
ensures that the occurrence of the Hawking effect is a gauge invariant statement \cite{di09}.

\section{Hawking effect}\label{che}

In this section, we develop the idea that the Hawking effect counteracts 
the classical direction of horizon crossing 
such that it can be applied to all types of trapping horizons.
We start with the analysis of future outer trapping 
horizons (FOTH) and 
recover Hawking's result via the Hamilton-Jacobi
method before we consider other horizon types.
Our treatment will be restricted to scalar
degrees of freedom but in principle 
higher spins would be possible. As an example, tunneling of 
fermions by dynamical black holes has been
analyzed with the outcome that the dominant contribution matches the results
for scalar fields  \cite{van08}. 

For a causal 
propagation, the equation of motion \eqref{kleingordon} will be
equipped with the Feynman prescription
$+i\varepsilon$, 
that is to say,
positive frequency modes propagate to the future and negative to the past.
Solutions to the Klein-Gordon equation with anti-Feynman prescription
$-i\varepsilon$ 
describe processes where positive frequency modes propagate to the past
and negative to the future, i.e. events are ordered complementary to the Feynman propagator.
Although we could perform all 
calculations with the Feynman prescription,
it will turn out that in some setups the anti-Feynman
propagator provides a clearer physical picture.

In the following subsections, we derive the Hamilton-Jacobi equation from 
 $(\Box-m^2/\hbar^2\pm i\varepsilon)\phi=0$ by applying it to 
\eqref{eis} and taking the semi-classical limit. Then,
the prescription in the 
Klein-Gordon equation will be transferred to the integral for $S_0$
as a complexification of the radial coordinate $R$ allowing the particle
to pass around the pole at the horizon.
Even though there is no classically allowed path across 
the horizon, quantum particles could escape through complex paths.

\subsection{Future outer trapping horizon}\label{foth}

Characteristic for future outer trapping horizons is that an
in-falling observer could easily enter but never escape.
It is illustrative to use the black hole as an example whenever 
we would like to clarify the physical idea behind our calculation
and compare the results with the existing literature.
A comprehensive analysis of dynamical black holes was performed
in \cite{vanzo}, where the special case of future outer trapping horizon was widely
analysed and several results for dynamical black holes were collected. 
As more explicit examples, the FOTHs ($\theta^+=0$)
of the Vaidya and the McVittie solution
have been studied in \cite{zoc07}. The McVittie metric describes a
static black hole in an expanding universe, hence, it has a future
outer black-hole horizon and a past inner cosmological horizon. By using the
cosmological expansion to give fiducial dynamics to the
black hole horizon, the Hawking
radiation for this specific dynamical FOTH is calculated via the tunneling method.

Nevertheless, our calculations account for spherically symmetric future outer 
trapping horizons. 
Consider a FOTH at radius $R_H$, 
the corresponding Hawking effect, as classically forbidden propagation,
is a tunnelling
from the trapped interior to the normal exterior region. 
In the Klein-Gordon equation, we use the Feynman prescription 
$+i\varepsilon$ because we treat a causal \textit{emission} 
towards the future.  We start to derive the 
relevant quantities along the Kodama vector
$K=(e^{-\psi(v,R)},\boldsymbol{0})$: the surface gravity
\begin{equation}\label{surf}
\kappa_H=\frac{1}{2}\partial_RC(R)|_{R=R_H},
\end{equation}
which has to be evaluated at the horizon and the Kodama energy 
for a particle in motion \cite{vanzo}
\begin{equation}\label{ene}
\omega=-K^\mu \partial_\mu S_0=-e^{-\psi(v,R)}\partial_vS_0.
\end{equation}
We could rewrite \eqref{phiwkb} by using \eqref{surf},\eqref{ene}, and
the identification with the momentum
 $k=\partial_RS_0$ to get
\begin{equation}
S_0=-\int e^{\psi(v,R)}\omega\mbox{d}v+\int k\mbox{d}R.
\end{equation}
Again, the angular parts are suppressed because of symmetry reasons.
Plugging this into the Hamilton-Jacobi equation with Feynman prescription, 
yields a quadratic equation for $k$
which can be solved to first order in $\varepsilon$. We get two
solutions
\begin{align}
k_1&=-\frac{i\varepsilon}{2\omega}\label{12},\\
k_2&=\frac{2\omega}{C(R)}+\frac{i\varepsilon}{2\omega}.\label{13}
\end{align}
These solutions represent the ingoing and outgoing directions
of motion. 
The momentum
$k_2$ describes solutions which are
going from inside to outside, while $k_1$ covers solutions which
are falling into the black hole.
We will see that only $k_2$ corresponds to a tunneling path,
because the roots of $C(R)$ cause this solution to have 
a pole in the radial coordinate, which can be bypassed on a complex path. 
Hence, $k_2$ generates an imaginary contribution
in the action
\begin{equation}\label{imsbh}
\mbox{Im}(S_0)=
\mbox{Im}\left(\int \frac{2\omega \mbox{d}R}
{C(R)\left(1-\frac{i\varepsilon C(R)}{4\omega^2}\right)}\right).
\end{equation}
Via a Taylor expansion of the function $C(R)$ around $R_H$ we can impose
a near horizon approximation
\begin{equation}\label{exo}
C(R)=(\partial_RC)|_{R=R_H}(R-R_H)+\mathcal{O}\left((R-R_H)^2\right).
\end{equation}
We would 
like to mention that $C(R)$ is
zero at $R_H$ \footnote{The function
$C(R)$ changes sign for $R>R_H$ and $R<R_H$. Therefore, and because $C(R)$
is a smooth function, the zero should be at the horizon.} 
while its derivative is finite and non-zero.
As already pointed out, the Hawking effect here is an emission and we could
describe it as a particle going from the trapped region into 
the normal region where it escapes to the observer. However, the definition
of particles and observers in the trapped region is a touchy business. Therefore,
we would like to use an equivalent picture such that we
could perform our experiment in the normal region, exploiting the features of the Kodama
vector.
Mathematically, an emission can be described similarly as \cite{par00}:
after pair-creation in the vicinity of the horizon, a
negative energy particle ($\omega\to-|\omega|$), or identically a hole, 
tunnels from the normal region
into the black hole interior. 
The resulting tunneling path of this process would be
the one of a negative energy particle traveling from outside 
to inside along a complexified path.
Together with \eqref{surf}, expansion \eqref{exo} can be 
plugged into \eqref{imsbh} yielding
\begin{equation}\label{imsbh1}
\mbox{Im}(S_0)=
\mbox{Im}\left(\int_{R_2}^{R_1} \frac{- |\omega| \mbox{d}R}
{\kappa_H\left(R-R_H-\frac{i\varepsilon'}{\kappa_H}\right)}\right),
\end{equation}
where $R_{1}<R_{H}<R_{2}$ and $\varepsilon'$ 
denotes the rescaled smallness parameter. Note, the $i\varepsilon$-prescription 
in the Klein-Gordan equation has become a complexification
of the radial coordinate.
After some manipulations 
\begin{equation}\label{imsbh2}
\mbox{Im}(S_0)=\lim_{\varepsilon''\to0}-\frac{|\omega|}{\kappa_H}
\int_{\delta/\varepsilon''}^{-\delta/\varepsilon''}\!\!\!\!\! \frac{1}
{\left(\frac{R-R_H}{\varepsilon''}\right)^2+1}
\mbox{d}\!\left(\frac{R-R_H}{\varepsilon''}\right),
\end{equation}
where we perform the integration
across the horizon with one coordinate in the trapped region and
one in the normal region such that $\delta=|R_1-R_2|$ is small
and we find
\begin{equation}\label{imsbh2}
\mbox{Im}(S_0)=\lim_{\varepsilon''\to0}-\frac{|\omega|}{\kappa_H}
\left[\arctan\left(-\frac{\delta}{\varepsilon''}\right)
-\arctan\left(\frac{\delta}{\varepsilon''}\right)\right].
\end{equation}
The integration yields an additional minus sign which would not have been present
for the positive energy particle tunneling outwards. However,
the final result will stay unaffected because of the sign change of $\omega$.
After taking the limit $\varepsilon''\to0$  
\begin{equation}\label{imp}
\mbox{Im}(S_0)=+\frac{\pi|\omega|}{\kappa_H}.
\end{equation}
The imaginary part is a positive number because
the surface gravity $\kappa_H>0$ for outer horizons. 
Equation \eqref{imp} can then be
used to calculate the tunneling rate \eqref{gamma}
\begin{equation}\label{gamma1}
\Gamma\propto \exp\left(-\frac{2\pi|\omega|}{\hbar\kappa_H}\right).
\end{equation}
As expected the Hawking effect for black holes obeys Im$(S_0)>0$ which mathematically 
reflects the classically prohibited process leading to an exponential suppression. To 
match the quantum mechanical tunneling rate and the temperature, we follow
the procedure outlined in \cite{hart83}.
The first step involves the Boltzmann distribution which gives
the probability to emit a particle, or in other words
we compare the probability of having $n$ particles to the
probability of having $n+1$ particles, at a fixed energy $E$
\begin{equation}\label{123}
P_{\rm emission}=
e^{-\tfrac{E}{k_BT}}P_{\rm absorption},
\end{equation}
with Boltzmann constant $k_B$.
It turns out that the emission probability
given by \eqref{123} is exponentially suppressed and can be associated
to the tunneling rate in \eqref{gamma1}. We can easily read off the corresponding parameter
\begin{equation}\label{th}
T=\frac{\hbar\kappa_H}{2\pi k_B},
\end{equation}
which coincides exactly with the temperature found by Hawking. We should note that
the relation between the quantity $T$ and the temperature of black hole
thermodynamics would have strictly applied if we had sent the observer to
the asymptotic regions, where other thermodynamic quantities, e.g.
the mass, are well-defined \cite{bek73}. Specifically, the parameter $T$ in the 
tunneling picture is generically a non-equilibrium temperature, since it is plagued by
gravitational modifications $T(t,R)=T_E/\Omega(t,R)+F(\Omega(t,R))$ with $T_E$ 
the equilibrium temperature and $F$ a complicated
function of the geometric factor $\Omega(t,R)$ (cf. \cite{fa14} for explicit calculations).
Considering metric \eqref{bardeen}, Hayward et. al. \cite{hay09} 
showed that for slowly evolving setups
$\Omega(t,R)\equiv\sqrt{C(R)}$ while $F(\Omega(t,R))$ becomes subdominant
with respect to the first term. 
Since the WKB approximation
requires the spacetime, and hence the horizon, to vary slowly, this constitutes just
a small deviation from the thermal equilibrium and approximates a full thermal 
equilibrium only in asymptotically flat regions.

Hence, the tunneling
description reproduces that an observer at  
future infinity measures a 
thermal spectrum coming from the FOTH.
From now on we will call the quantity $T$ (Hayward-Kodama) temperature with the warning
that this identification is only valid in an appropriate approximation such as 
in the asymptotic regions of spacetime \cite{bach99,par04}.

\subsection{Future inner trapping horizon}\label{fith}

The following subsection shows that for the future inner
trapping horizon a different sort of Hawking effect exists. In particular, we will 
verify that 
the FITH is the absorptive partner of the FOTH with respect to quantum effects.
Future inner trapping horizons appear, for example, in contracting cosmologies with 
a big crunch in the future, as well as 
in more general black hole solutions, like the Reissner-Nordstr\"om black hole, in
combination with a FOTH.
Hence, to understand the Hawking effect for general
black hole solutions it is, as a first step, important 
to understand the Hawking effect for FITHs.

Before we explore the actual calculation, we analyze the 
setup from the tunneling perspective.
FITHs enclose the
normal region inside while the trapped region lies beyond the 
horizon. In the example of a big
crunch, the universe contracts such that, at a distance $R_H$ to the
observer, the contraction reaches apparently
the speed of light. Classically, all signals sent by the observer
at $R=0$ will be bound to distances
$R\le R_H$, i.e. the classically forbidden path would be to cross the FITH
from the normal into the trapped region. 
As quantum effect, we therefore expect to see an \textit{absorption} of particles by the horizon;
referring to our gedankenexperiment, 
the observer would count less particles in the final than in the initial state.
In the normal region, we could portray an absorption in two ways: either we describe
the system by a causal Feynman propagation and allow the horizon
to emit negative energy particles/holes;
or alternatively
we illustrate this process through an emission
of positive energy particles by the horizon towards the past. 
In the first case, the negative energy particles emitted by
the horizon annihilate some of the initially prepared particles while the 
latter case could be regarded as adding particles to the initial state
which then get absorbed during a causal
time evolution. In both descriptions, the net process is a detection
of less particles in the observer's final state.

To connect FITHs to thermodynamics,
the easiest way would be to choose the second description using the anti-Feynman prescription,
i.e. $-i\varepsilon$, and count particles emitted towards the past. Then, we observe
a spectrum which we can directly
relate to the temperature of the horizon. 
In the tunneling picture, emission into the past coincides with a Hawking pair where the
negative energy particle crosses the horizon
while the positive energy particle reaches the observer's initial state,
traveling both backwards in time.
According to this, we analyze
for the FITH scenario negative energy particles which
path goes from the normal region inside ($R_1<R_H$) into 
the trapped region outside ($R_2>R_H$) but backwards in time.
The Hamilton-Jacobi equation yields similar
solutions for $k$ as in \ref{foth}
\begin{align}
k_1&=+\frac{ i\varepsilon}{2\omega},\\
k_2&=\frac{2\omega}{C(R)}-\frac{i\varepsilon}{2\omega}.
\end{align}
but now with the anti-Feynman prescription.
Again, $k_2$ admits a pole at the horizon leading to a non-zero
tunneling probability. 
With this in mind, we get as imaginary part of $S_0$
\begin{equation}
\mbox{Im}(S_0)=
\mbox{Im}\left(\int_{R_1}^{R_2}\frac{-|\omega| \mbox{d}R}
{-|\kappa_H|\left(R-R_H+\frac{i\varepsilon'}{-|\kappa_H|}\right)}\right).
\end{equation} 
Due to the fact that $\kappa_H<0$ for inner horizons, the imaginary part
acquires the same value as in \ref{foth}. This results in a non-zero tunneling rate
\begin{equation}
\Gamma\propto \exp\left(-\frac{2\pi|\omega|}{\hbar|\kappa_H|}\right)
\end{equation}
showing that there is a non-zero probability for the particle to get absorbed by 
the horizon. Comparison with the Boltzmann distribution \eqref{123} yields as
temperature of the emitted spectrum
\begin{equation}\label{tenp}
T=\frac{\hbar|\kappa_H|}{2\pi k_B}.
\end{equation}
With the tunneling picture we could transfer our principle
idea about the Hawking effect and find that FITHs 
are subjected to an absorptive Hawking effect. With respect to 
the definition, FITHs are the absorptive partners to the emissive FOTHs. 

Note, we could have calculated everything 
in the Feynman prescription; in that case we
would have to look at the correct process 
and change \eqref{123} accordingly to be
\begin{equation}\label{321}
P_{\rm absorption}=
e^{-\tfrac{E}{k_BT}}P_{\rm emission}
\end{equation}
because now the less probable process is given by the absorption
of a positive energy particle.  For an emission of a negative energy 
particle we have to invert \eqref{321} but since we compare negative
energy emission, we need to replace $E\to -|E|$. As we mentioned before
we compare the tunneling rate with the probability \eqref{321} of going from
$n$ positive energy particles to $n-1$, or equivalently from 
$n$ negative energy particles to $n+1$, resulting in a positive 
temperature like \eqref{tenp}. In this
light, the formalism gives a consistent result and supports
the proposed interpretation.
Nevertheless, one could worry about measuring negative 
energies in the system, but the physical interpretation is
still consistent because the emitted negative energy particles 
deplete the initially prepared state and lower the number of particles
in the normal region. In order to be more illustrative, we
could exploit the big crunch a bit more considering now a photon test field 
in our contracting universe. We would like to do the same gedankenexperiment
as before and measure particles arriving at our detector. 
In this cosmology, an observer
would experience a constant influx through the horizon caused by the 
presence of the photons. This is predicted
by \eqref{321} telling us that the emission process is geometrically
favored while the absorption is exponentially suppressed. 
If we now switch on quantum effects, we will observe a 
reduction of the natural emission spectrum by the absorption
spectrum associated to $T$. 

\subsection{Past inner trapping horizon}\label{pith}

In the next two subsections, we show that an emissive/absorptive
horizon pair exists also for past horizons.
Let us take an expanding cosmology with 
Big Bang as example where the universe expands with a Hubble law
from the initial singularity.
Since the expansion is homogeneous and isotropic,
the observer at $R=0$ can only see events inside the Hubble sphere with radius $R<R_H$ and
will not be able to see beyond
because all events at $R>R_{H}$ are trapped outside the horizon \cite{dav04}.
For Friedmann-Lema\^itre-Robertson-Walker spacetimes, which admit 
a PITH, the Hawking effect has been investigated in
\cite{vanzo} and found, similar to the black hole case, to be
an \textit{emission} with a thermal spectrum.

The Hamilton-Jacobi formalism requires to identify
the classically forbidden process which is now 
an emission of particles from the anti-trapped region outside the
Hubble sphere into the normal region
inside. Again, this is a causal emission into the future
and we adopt the Feynman prescription $+i\varepsilon$ in the 
Klein-Gordon equation. In the tunneling picture this could be
recasted as a negative energy particle crossing the horizon 
from the normal region at $R_1<R_{H}$ to $R_2>R_{H}$.
Therefore, we can perform the steps similar to section \ref{foth} and
solve the Klein-Gordon equation 
with a Feynman prescription. Whenever we treat a past trapping horizon,
we use \eqref{bardeen} with outgoing lightcone coordinate $\xi=u$.
Altogether, the Hamilton-Jacobi equation yields two different
solutions for the momentum
\begin{align}
k_1&=-\frac{i\varepsilon}{2\omega}\label{28},\\
k_2&=-\frac{2\omega}{C(R)}+\frac{i\varepsilon}{2\omega}\label{29}.
\end{align}
with $k_1$ now being the outgoing and $k_2$ the ingoing solution
to first order in $\varepsilon$. Carrying out the same procedure as in the previous
sections we find for the negative energy particle
\begin{equation}\label{imsbh1}
\mbox{Im}(S_0)=
\mbox{Im}\left(\int_{R_1}^{R_2} \frac{- |\omega| \mbox{d}R}
{-|\kappa_H|\left(R-R_H-\frac{i\varepsilon'}{-|\kappa_H|}\right)}\right)
\end{equation}
where the surface 
gravity is now negative 
$\kappa_H\to-|\kappa_H|$.
As before, the non-zero imaginary part appears because of a pole
at the horizon and we could 
interpret the resulting process as a net particle flux into the
observer's Hubble sphere.
The imaginary part equals 
\eqref{imp}, the resulting rate equals \eqref{gamma1}, and the emitted 
thermal spectrum peaks at a temperature
\begin{equation}
T=\frac{\hbar|\kappa_H|}{2\pi k_B}.
\end{equation}
Note that this is in perfect agreement with the temperatures found for PITHs in an FLRW 
\cite{vanzo, cai09,far11} as well as in a de Sitter background \cite{gibb77}.

The two horizons, PITH and FOTH, are both subjected to a Hawking effect
which is described by an emission of particles (towards the future). Therefore, 
from the perspective of quantum field theory both
horizons are emissive horizons. 

\subsection{Past outer trapping horizon}\label{poth}

Past outer trapping horizons can occur in a fully extended black-hole spacetime
describing white hole horizons or appear in
cosmologies dominated by a stiff fluid, i.e. equation of state parameter $w=1$. 
The latter example has both a past inner and a past outer trapping horizon.
For the sake of 
developing intuition, we will consider the white hole as the main example.
Since white holes can classically just emit, the Hawking 
effect, as counter-effect, is expected to act as an \textit{absorption}.
By definition, POTHs enclose the trapped region inside at $R<R_H$ while the normal region
is located outside.
We want to specify that the trapped region 
is, in fact, anti-trapped, i.e. classical trajectories depart from each other 
in the future development such that the geometrically favored direction is outgoing. 
Then, the process of interest is the 
path entering the white hole or, in other words,
traverses the POTH from the normal into the
anti-trapped region. 
Following the arguments in \ref{fith}, we employ the anti-Feynman prescription to 
describe absorption, that is, we prepare the initial state  
in the future and perform a backward-in-time measurement. To assign a temperature to this 
emitted spectrum, we 
collect the particles which are emitted towards the past.
The calculations in section \ref{foth} - \ref{pith} change as follows: the
solutions for $k$ are given by \eqref{28} and \eqref{29} but with 
the anti-Feynman prescription
\begin{align}
k_1&=+\frac{i\varepsilon}{2\omega},\\
k_2&=-\frac{2\omega}{C(R)}-\frac{i\varepsilon}{2\omega}.
\end{align}
Again, only the ingoing momentum
$k_2$ contributes to the tunneling rate because of the pole induced by the zero in $C(R)$.
The corresponding imaginary part reads
\begin{equation}\label{imswh12}
\mbox{Im}(S_0)=
\mbox{Im}\left(\int_{R_2}^{R_1}\frac{-|\omega| \mbox{d}R}
{\kappa_H\left(R-R_H+\frac{i\varepsilon'}{\kappa_H}\right)}\right)
\end{equation}
where $\kappa_H$ is now positive. 
The above integral \eqref{imswh12} describes a negative energy particle
tunneling backwards in time into the anti-trapped region. This accounts
for a net particle flux which the horizon emits towards the  past \cite{par00}.
The resulting imaginary part
\begin{equation}\label{imswh1}
\mbox{Im}(S_0)=+
\frac{\pi|\omega|}{\kappa_H}
\end{equation}
induces a non-zero tunneling rate. In other words, the non-removable 
pole at $R_H$ furnishes the POTH with an absorptive Hawking effect.
We see again that the anti-Feynman prescription leads to a thermal 
spectrum with temperature
\begin{equation}
T=\frac{\hbar\kappa_H}{2\pi k_B},
\end{equation}
emitted towards the past.
When we consider the example of an emitting white hole, we would 
conclude that the Hawking effect reduces the spectrum 
seen by a future observer, who could interpret this as 
emission of holes or an absorption of particles, like in section \ref{fith}. POTHs are
therefore the absorptive partner of PITHs and our results for POTHs
are in accordance to the idea of \cite{vanzo} about the white hole event horizon,
which is a static null version of POTHs.

\section{Discussion}

Hayward classified trapping horizons with respect to null 
expansions along in- and outgoing light rays into four types. 
In this article, we analyzed quantum effects associated with these horizons and
verified for spherically symmetric spacetimes that all these trapping horizons are
subjected to a Hawking effect by applying
the tunneling picture via the Hamilton-Jacobi method. 
The idea is to formulate the 
Hawking effect as the propagation opposite to the
path preferred by
general relativity. In the tunneling picture, the horizon
might induce a pole in the action which excludes classical
paths. However, a remedy can be found by looking at
quantum processes: the consistency of the theory stays unharmed as long as
avoiding poles on complexified paths happens
on time scales that do not violate macrocausality.
According to this, we found that past outer and future inner trapping 
horizons are subjected to an absorptive Hawking effect.

Within our framework, quantum field theory on curved spacetime suggests consequently two types of horizons:
absorptive horizons (FITH and POTH) for which the Hawking effect lowers the number of
detected particles and emissive horizons (FOTH and PITH) for which radiation from the
horizon occurs. To strengthen the physical intuition related to the
absorption effect, we consider a 
black hole which admits a future inner and
future outer trapping horizon together with a suitably stable inner normal region \footnote{
Recent publications show that the innermost normal region of Reissner-Nordstr\"om black holes
are highly unstable under small perturbations \cite{curiel1} while in Kerr black holes this 
region is stable \cite{curiel2}. Due to the axisymmetry this needs further investigation from the perspective of
the tunneling method.}.
By energy budget considerations
the Hawking effect induces an evaporation of black holes. Reducing the black hole mass implies
to extract energy out of the black hole interior. 
Tunneling paths for 
black holes admitting an inner and outer horizon have therefore to cross both.
According to our gedankenexperiment,
an observer in the interior, surrounded by the inner horizon,
i.e. in the normal region, will prepare particles and send them 
towards the horizon. After some time has elapsed, the observer
counts the particles
and will measure a depletion of particles because of the Hawking absorption.
From this, the observer would infer some particles have crossed
the future inner horizon into the trapped 
region. To get the energy finally out of the black hole, the particles have to 
cross the outer horizon as well. An outside observer would then perceive an
ordinary Hawking emission.
From this perspective, the result that one realization of the 
Hawking effect corresponds to an absorption seems reasonable. 
Because the probability for 
traveling fully through the trapped region is
strongly restricted by the evanescence of the wave-function,
we could estimate the lifespan of black holes with two horizons
compared to Schwarzschild black holes 
from the tunneling perspective of Hawking radiation. 
Intuitively, we hypothesize that crossing two trapping horizons 
through the trapped region reduces the 
overall tunneling probability which will
enhance the life-time
accordingly.
Additionally, charged black holes suffer from the problem of mass inflation \cite{poi89}. 
A full consideration of the Hawking effect might be helpful to address this 
issue. However, this scenario
needs further investigations in the future.

On the other hand, our results for FITHs and POTHs need further discussions.
Considering a radiation filled contracting
cosmology or white hole, an absorptive Hawking effect is the
equivalent of a thermal emission of holes which reduce the ejection caused by geometry. 
However, there is 
an important difference between emissive and absorptive Hawking effect:
in case of FOTHs and PITHs
the horizon crossing occurs from the trapped (or anti-trapped respectively) 
to the normal region while for FITHs and POTHs the Hawking effect describes 
an absorption into the trapped (or anti-trapped) region. 
The normal region supports particle to stay there, but the trapped (or anti-trapped) region
of FITHs and POTHs will force the particles to leave as soon as possible. Hence, 
the Hawking process gets revoked. While this
takes place, already new Hawking particles will have entered the (anti-)trapped
region, possibly slowing down the cosmic contraction in
big crunch scenarios or increasing a white hole's lifetime. The observable contribution from the 
Hawking effect will not result in the abundance but in the absence
of particles. This raises the question of whether a Hawking 
effect for FITHs and POTHs occurs as an in-principle 
detectable thermal spectrum. We doubt that this is the case
because the released particle might recombine with its Hawking partner
in the normal region and the effective particle number
stays unchanged. The only detectable thermal spectrum will occur in the 
case when the absorbing horizons are cascaded with emitting  
horizons. A penetration of FOTHs and PITHs is likely but a total, eternal absorption seems to 
be unlikely unless there is a horizon providing an additional exit from the trapped region.

Possible applications of these results exist 
in gravitational collapse scenarios which lead to the 
formation of black holes. Inclusion of the Hawking effect could
regularize the collapse such that trapped surfaces
similar to black holes will form but without ever
forming a singularity \cite{hay06,roman,bine18}.
In some classes of collapsing solutions such as the Vaidya spacetime,
a future inner and a future outer trapping horizon are present. 
The evolution of the inner trapping horizon, predicted by general relativity,
finally forms a singularity while 
the outer asymptotically approaches
the black hole event horizon. In this case, the Hawking
effect leads the outer horizon to shrink and 
our results predict a slowdown in the evolution of the inner
horizon such that both horizons could approach each other.
 In particular, the possibility for the occurrence of
a closed trapped surface without the formation of a singularity opens up when both
horizons meet. In this sense, the Hawking effect could facilitate a stellar 
collapse without singularity.

Sebastiani, Vanzo, and Zerbini \cite{zerb18} showed that traversable static
wormhole horizons are not subjected to a Hawking process. Their result
does not raise a contradiction to ours, instead, it fits into 
the presented physical idea: the wormholes in \cite{zerb18} are in principle traversable
and an observer could always enter and exit. Such static wormhole horizon
lacks the property of being semipermeable. Mathematically, this is explained by 
the presence of a removable  
pole at the horizon with the consequence that the action has no imaginary 
part causing a Hawking effect. 

Finally, it should be noticed that our argumentation might be generalisable beyond
spherically symmetric spacetimes. 
In \cite{seno15}, a generalisation of the Kodama vector has been proposed. This dual expansion
vector, for which similar quantities as the energy along the flow as well as the surface gravity 
can be defined, serves as a possible path towards an application of the presented formalism to
more general spacetimes. Since only FOTHs and PITHs have been studied by the authors, its 
application to FITHs and POTHs and the possibility of a Hawking absorption 
needs to be investigated in the future.

\ack
It is a great pleasure to thank Richard Bond, David Brizuela,
Alexis Helou, Maximilian K\"ogler, Stefan Hofmann,
Fr\'ed\'eric Lamy, and Ue-Li Pen
for delightful discussions and their thoughts on the topic. We are especially grateful to
Sergio Zerbini for his comments and thoughts 
on the manuscript and his great eagerness in discussing our ideas. 
The authors appreciate financial support 
from the Cusanuswerk e.V. and the Alexander von Humboldt-Stiftung. 
\section*{References}
\bibliographystyle{iopart-num.bst}
\bibliography{FIHLit.bib}

\end{document}